# Instrument for Measuring the Earth's Time-Retarded Transverse Gravitational Vector Potential

**J. C. Hafele**[1]

Here within the basic design for a ground-based instrument for measuring the magnitude of the Earth's time-retarded transverse gravitational vector potential is described. The formula for the Earth's transverse vector potential is derived from the known formula for the neoclassical time-retarded transverse gravitational field (arXiv:0904.0383v2 [physics.gen-ph] 25May2010). The device senses the relativistic shift in the frequency of laser-diode oscillators set into circular motion at the tips of a two-arm rotor. The instrument employs fiber optics and a digital electronic interferometer/spectrometer to measure the effect of the relativistic time dilation on the frequency-modulated (FM) harmonic amplitudes in the beat signals between the tip-diodes and a stationary reference diode. The FM amplitudes depend on the orientation of the rotor. For the vertical-east-west orientation with a rotor frequency of 73.9 Hz, the predicted FM amplitudes for overtones at 148 Hz, 222 Hz, and 296 Hz are respectively $7\times10^{-10}$ Hz, $4\times10^{-11}$ Hz, and $9\times10^{-11}$ Hz. The overtones in the beat signals can be amplified and observed with a tunable FM digital audio amplifier. The measured values for the harmonics of the vector potential can be determined by back-calculating what the amplitudes must have been at the input to the amplifier. The instrument can be used to establish the speed of the Earth's gravitational field and to study the structure of the Earth's mantle and outer core.

## 1. INTRODUCTION

The mystery of the flyby anomaly has been resolved. In May of 2010, this author published a new neoclassical time-retarded theory that explains exactly the small anomalous speed change reported in 2008 by NASA scientists for six spacecraft flybys of the Earth.[(1)]

The neoclassical time-retarded theory is based on the slow-speed weak-field approximation for general relativity theory.[(2)] Any causal interaction requires time retardation, which means that it takes a certain amount of time for a causal field to propagate from a moving point-mass source to a possibly moving distant field-point. The point-mass sources in a large rotating sphere (a simulation of the Earth) are in circular motion around the rotational axis of the sphere. This circular motion generates a time-retarded transverse gravitational field that is orthogonal to the classical Newtonian radial gravitational field. The transverse field is proportional to the ratio of the equatorial surface speed of the rotating sphere to the speed of propagation of the gravitational field. Motion of the field point may cause a time-dependent transverse gravitational field. This time dependence generates an induction field that is proportional to the time derivative of the transverse field. The predicted speed-change for a flyby is proportional to the induction constant.

[1] 618 S. 24th St., Laramie, WY 82070, USA, e-mail: cahafele@bresnan.net

The exact numerical value for the induction constant is at this time uncertain, but the observed speed changes for six Earth flybys indicate that the order-of-magnitude is $10^{-4}$ s/m. The objective of this report is to describe a feasible design for a ground-based instrument that can produce an exact measure for the induction constant, and thereby give a direct measure for the speed of the Earth's gravitational field.

The proposed instrument employs a high-speed two-arm rotor with laser diodes at the tips. If a digital interferometer/spectrometer is used to superimpose the laser beams, the resulting harmonics in the beat signals would be frequency-modulated (FM). The beat-signals for the vertical-east-west orientation would contain detectable overtones from the first to the third. Order-of-magnitude estimates for the FM harmonic amplitudes for the vertical-east-west orientation are listed in Table I (page 23).

In section 2.1, the dependences of the various gravitational acceleration fields on u/v for a large rotating sphere in the slow-speed weak-field approximation are reviewed. The classical radial acceleration field is independent of (u/c), and the neoclassical transverse acceleration field depends on $(u/c)^1$. The gravitomagnetic fields, being of higher order in (u/c), are negligible.

There are three relevant potentials; the scalar potential for the classical Newtonian radial gravitational field, a vector potential for the neoclassical time-retarded transverse gravitational field, and another vector potential for the induction field. Formulas for the needed potential functions are derived in section 2.2.

The relativistic proper time recorded by a clock depends on the speed and on the three potential functions. The formula for the proper time for clocks near the Earth's surface, in the slow-speed weak-field approximation, is derived in section 2.3.

The formula for the relativistic proper frequency for a clock's oscillator, which depends on the clock's proper time, is derived in section 3.1. The proper frequencies for laser-diode oscillators in a two-arm rotor depend on the rotor's orientation, which means the resulting beat signals also depend on the rotor's orientation. Order-of-magnitude estimates for the FM harmonics in the vertical-east-west orientation, the vertical-north-south orientation, and the horizontal-east-west orientation, are derived in sections 3.2, 3.3, and 3.4. The vertical-east-west orientation is the only one that produces a rich array of detectable harmonics. The other orientations can be used to test the instrument.

## 2. GRAVITATIONAL FIELDS OF A LARGE ROTATING SPHERE IN THE SLOW-SPEED WEAK-FIELD APPROXIMATION

### 2.1 Orders of u/c for the Gravitational Fields of a Large Rotating Central Sphere

Assume the Earth can be simulated by a large rotating isotropic sphere of radius $r_E$, mass $M_E$, angular speed $\Omega_E$, and radial mass-density distribution $\rho_E$. In the slow-speed weak-field approximation, general relativity theory reduces to classical time-retarded electromagnetic field theory. Formulas for the gravitational fields in the slow-speed weak-field approximation can be found in W. Rindler's textbook.[2] The formula that is analogous to the "Lorentz Force Law" is

$$\mathbf{a} = -\mathbf{e} - \frac{1}{c}\mathbf{u} \times \mathbf{h} \ , \qquad (2.1.1)$$

where **a** is the gravitational acceleration field of the central sphere, **e** is the gravitational equivalent of the electric field, **h** is the gravitational equivalent of the magnetic induction field, **u** is the velocity of a field point outside of the central sphere in a nonrotating (inertial) geocentric frame of reference, and c is the vacuum speed of light. The field **e** is defined by the gradient of the gravitational scalar potential, **grad**$\chi$, and the field **h** is defined by the curl of the gravitational vector potential, **curlV**.

Rindler defines the time-retarded potential functions for **e** and **h** by using a cryptic notation, $[\rho]$ and $[\rho\mathbf{u}]$, to denote "the value 'retarded' by the light travel time to the origin of r," where r is the radial distance from a circulating point-mass of mass $\rho dV$ to the field point.[2] The formulas for $\chi$ and **V** using this notation are

$$\chi = -G\iiint \frac{[\rho]\,dV}{r} \ , \qquad \mathbf{V} = \frac{1}{c}G\iiint \frac{[\rho\mathbf{u}]\,dV}{r} \ . \qquad (2.1.2)$$

Let's clarify what this notation means for a large central rotating isotropic sphere. Let $\rho_E$ be the radial mass-density distribution, and let $\bar{\rho}_E$ be the mean value for $\rho_E$. The formulas for $\chi$ and **V** can be rewritten as

$$\chi = -G\bar{\rho}_E\iiint \frac{\left[\rho_E/\bar{\rho}_E\right]dV}{r} \ , \qquad \mathbf{V} = \frac{1}{c}G\bar{\rho}_E\iiint \frac{\left[\mathbf{u}\,\rho_E/\bar{\rho}_E\right]dV}{r} \ . \qquad (2.1.3)$$

Let t be the "coordinated time" at the field point and let t′ be the "retarded time" at the circulating source point. Let the space and time coordinates for the field point be $(r,\lambda,\phi,t)$, where the definition for $\mathbf{r}(t)$ (bold face Courier) is now changed to be the radial vector from the center of the sphere to the field point. Let the space and time coordinates for the source point be ($\mathsf{r}$,λ′,ϕ′,t′), where $\mathbf{r}$(t′) (bold face Arial) is the radial vector from the center of the sphere to the source point. The space variables r,λ,ϕ are functions of t, and the space variables $\mathsf{r}$,λ′,ϕ′ are functions of t′.

To satisfy causality, a gravitational signal emitted at t′ by a point-mass at the source point arrives at the field point at a slightly later time t. Let $\mathbf{r}'$ be the "retarded vector distance" from the circulating source-point to the field-point, defined by $\mathbf{r}'(t,t')\equiv\mathbf{r}(t)-\mathbf{r}(t')$. If the speed of the signal is c, and r′ is the absolute magnitude of $\mathbf{r}'$, the formula relating t to t′ is

$$t = t' + \frac{r'}{c} \quad . \tag{2.1.4}$$

The retarded-time derivative, dt/dt′, is

$$\frac{dt}{dt'} = 1 + \frac{1}{c}\frac{dr'}{dt'} \quad . \tag{2.1.5}$$

The formulas for χ and $\mathbf{V}$ (Eq. (2.1.3)) can be rewritten explicitly in terms of r′(t,t′) and $dV(t')=\mathsf{r}^2\cos(\lambda')\,d\mathsf{r}\,d\lambda'\,d\phi'$, as follows.

$$\chi = -G\bar{\rho}_E r_E^{\,3}\int_0^{r_E}\frac{\rho_E}{\bar{\rho}_E}\frac{\mathsf{r}^2}{r_E^{\,2}}\left(\int_{-\pi/2}^{\pi/2}\cos(\lambda')\left(\int_{-\pi}^{\pi}\frac{d\phi'}{|\mathbf{r}(t)-\mathbf{r}(t')|}\right)d\lambda'\right)\frac{d\mathsf{r}}{r_E} \quad ,$$

$$\mathbf{V} = \frac{G\bar{\rho}_E r_E^{\,3}}{c}\int_0^{r_E}\frac{\rho_E}{\bar{\rho}_E}\frac{\mathsf{r}^2}{r_E^{\,2}}\left(\int_{-\pi/2}^{\pi/2}\cos(\lambda')\left(\int_{-\pi}^{\pi}\frac{\mathbf{u}(t)d\phi'}{|\mathbf{r}(t)-\mathbf{r}(t')|}\right)d\lambda'\right)\frac{d\mathsf{r}}{r_E} \quad . \tag{2.1.6}$$

After the triple integration over the volume of the sphere, the functions χ and $\mathbf{V}$ depend only on the time t at the field point. The "driver" time t′ is pre-determined by the rotation of the central sphere. The "driven" time t is post-determined by the motion of the field point. The motion of the field point may or may not be constrained by nongravitational forces.

The integrand for the triple integral over the volume is a cyclical function of t′, and t′ is in turn a cyclical function of t. The dependence of χ on t can be found by integration over t from 0 to t. But first there is the triple integration over the volume of the sphere. This first integration can be accomplished by transforming the integration variable from t to t′. By using the Jacobian for this transformation, dt/dt′, the integration over time results in two integrals, as follows.

$$\int_{0}^{t} \chi(t)dt = \int_{t'(0)}^{t'(t)} \chi(t')\frac{dt}{dt'}dt' = \int_{t'(0)}^{t'(t)} \chi(t')dt' + \int_{t'(0)}^{t'(t)} \chi(t')\frac{1}{c}\frac{dr'}{dt'}dt' \quad . \qquad (2.1.7)$$

The first integral, call it $\chi_0(t)$, gives the classical Newtonian instantaneous action-at-a-distance scalar potential for **e**. The second integral, call it $\chi_1(t)$, gives the "time-retarded" potential function that is proportional to the ratio (dr'/dt')/c. The second integral gives the potential function for a vortex field. Therefore, the gradient, **grad**$\chi_1(t)$, is zero, but the curl, **curl**$\boldsymbol{\chi}_1(t)$, is not zero, which means the function $\boldsymbol{\chi}_1(t)$ (bold face chi) is a <u>vector</u> potential function. The formula for the total gravitational potential for **e**, where $\chi_1(t)$ is the magnitude of the vector $\boldsymbol{\chi}_1(t)$, becomes

$$\chi_0(t) = \int_{t'(0)}^{t'(t)} \chi(t')dt' \quad ,$$

$$\chi_1(t) = \int_{t'(0)}^{t'(t)} \chi(t')\frac{1}{c}\frac{dr'}{dt'}dt' \quad ,$$

$$\boldsymbol{\chi}(t) = \chi_0(t) + \boldsymbol{\chi}_1(t) \quad . \qquad (2.1.8)$$

The formulas for the scalar potential $\chi_0$ and the magnitude of the vector potential $\boldsymbol{\chi}_1$ become (cf. Eqs. (2.1.2) and (2.1.6))

$$\chi_0 = -G\iiint \frac{\rho_E dV}{r'} \quad ,$$

$$\chi_1 = G\iiint \frac{1}{c}\frac{dr'}{dt'}\frac{\rho_E dV}{r'} \quad , \qquad (2.1.9)$$

where the triple integral is defined by Eq. (2.1.6). Notice that $\chi_0$ is of order $(u/c)^0$ and $\chi_1$ is of order $(u/c)^1$.

The formula for the analogous gravitational electric field becomes (cf. Eq. (2.1.1))

$$\mathbf{e} = \mathbf{e}_0 + \mathbf{e}_1 = -\mathbf{grad}\chi_0 - \mathbf{curl}\boldsymbol{\chi}_1 \quad . \qquad (2.1.10)$$

Next consider the vector potential **V** for **h**. The instantaneous and time-retarded integrals for **V** become

$$\int_{0}^{t} \mathbf{V}(t)dt = \int_{t'(0)}^{t'(t)} \mathbf{V}(t')\frac{dt}{dt'}dt' = \int_{t'(0)}^{t'(t)} \mathbf{V}(t')dt' + \int_{t'(0)}^{t'(t)} \mathbf{V}(t')\frac{1}{c}\frac{dr'}{dt'}dt' \quad ,$$

$$\mathbf{V}_0(t) = \int_{t'(0)}^{t'(t)} \mathbf{V}(t')dt' \quad ,$$

$$\mathbf{V}_1(t) = \int_{t'(0)}^{t'(t)} \mathbf{V}(t')\frac{1}{c}\frac{dr'}{dt'}dt' \quad . \tag{2.1.11}$$

The formulas for $\mathbf{V}_0$ and $\mathbf{V}_1$ become

$$\mathbf{V}_0 = G\iiint \frac{\mathbf{u}}{c}\frac{\rho_E dV}{r'} \quad ,$$

$$\mathbf{V}_1 = G\iiint \frac{\mathbf{u}}{c}\frac{1}{c}\frac{dr'}{dt'}\frac{\rho_E dV}{r'} \quad . \tag{2.1.12}$$

Notice that $\mathbf{V}_0$ is of order $(u/c)^1$ and $\mathbf{V}_1$ is of order $(u/c)^2$.

The formula for the analogous gravitational magnetic induction field (gravitomagnetic fields) becomes[2]

$$\mathbf{h} = \mathbf{h}_0 + \mathbf{h}_1 = 4\mathbf{curl}\mathbf{V}_0 + 4\mathbf{curl}\mathbf{V}_1 \quad . \tag{2.1.13}$$

Therefore, the orders for the analogous Lorentz Force Law become (cf. Eq. (2.1.1))

$$\mathbf{a}_0 = -\mathbf{e}_0 = -\mathbf{grad}\chi_0 \ , \text{ order } (u/c)^0 \ ,$$
$$\mathbf{a}_1 = -\mathbf{e}_1 = -\mathbf{curl}\mathbf{X}_1 \ , \text{ order } (u/c)^1 \ ,$$
$$\mathbf{a}_2 = -\frac{4}{c}\mathbf{u} \times \mathbf{h}_0 = -\frac{4}{c}\mathbf{u} \times \mathbf{curl}\mathbf{V}_0 \ , \text{ order } (u/c)^2 \ ,$$
$$\mathbf{a}_3 = -\frac{4}{c}\mathbf{u} \times \mathbf{h}_1 = -\frac{4}{c}\mathbf{u} \times \mathbf{curl}\mathbf{V}_1 \ , \text{ order } (u/c)^3 \ . \tag{2.1.14}$$

This shows that the analogous gravitational magnetic induction field, relative to $\mathbf{X}_1$, is increased slightly by the "novel" factor of 4, but is decreased by the huge denominators of c or $c^2$. The acceleration field, $\mathbf{a}_0$, is the classical Newtonian instantaneous action-at-a-distance radial gravitational field, $\mathbf{g}_r$, where the subscript "r" indicates that this field is directed radially outward or inward. The acceleration field, $\mathbf{a}_1$, is the neoclassical time-retarded transverse gravitational field, $\mathbf{g}_e$, where the subscript "e" indicates that this field is directed towards the east or west. The remaining acceleration fields are of higher order in (u/c), and can be neglected for a first order approximation.

The neoclassical transverse gravitational field, $\mathbf{g}_e$, causes a small perturbation on the classical Newtonian field, $\mathbf{g}_r$. It has been shown that an induction field, designated $\mathbf{F}_\lambda$, that is proportional to the time rate of change of $\mathbf{g}_e$, explains exactly the observed anomalous change in the speed of six Earth spacecraft flybys reported by NASA scientists.[1]

## 2.2 Gravitational Potential Functions for a Large Rotating Sphere in the Slow-Speed Weak-Field Approximation

Let $\mathbf{e}_r$, $\mathbf{e}_\lambda$, and $\mathbf{e}_\phi$ be orthogonal unit vectors for the space-time spherical system $(r,\lambda,\phi,t)$; $\mathbf{e}_r$ is directed radially outward, $\mathbf{e}_\lambda$ is directed southward, and $\mathbf{e}_\phi$ is directed eastward. The polar angle or colatitude increases towards $+\mathbf{e}_\lambda$ (southward); the latitude $\lambda$ increases towards $-\mathbf{e}_\lambda$ (northward). The dot products $\mathbf{e}_r \bullet \mathbf{e}_\lambda=0$, etc. The cross products $\mathbf{e}_r \times \mathbf{e}_\lambda=\mathbf{e}_\phi$, etc. The self dot products $\mathbf{e}_r \bullet \mathbf{e}_r=1$, etc. The self cross products $\mathbf{e}_r \times \mathbf{e}_r=\mathbf{0}$, etc. The time t is the observed coordinated time at the field-point.

The formula for the "proper" velocity of the field point, $\mathbf{u}$ in the $(r,\lambda,\phi,t)$ space-time system, becomes

$$\mathbf{u} = \mathbf{e}_r u_r + \mathbf{e}_\lambda u_\lambda + \mathbf{e}_\phi u_\phi = \mathbf{e}_r \frac{dr}{dt} - \mathbf{e}_\lambda r_\lambda \frac{d\lambda}{dt} + \mathbf{e}_\phi r_\phi \frac{d\phi}{dt} \quad . \qquad (2.2.1)$$

The formula for the gravitational field $\mathbf{g}$ in the $(r,\lambda,\phi,t)$ system becomes

$$\mathbf{g} = \mathbf{g}_r + \mathbf{g}_e = \mathbf{e}_r g_r + \mathbf{e}_\phi g_e \quad . \qquad (2.2.2)$$

The transverse gravitational field is perpendicular to the radial gravitational field, $\mathbf{g}_r \bullet \mathbf{g}_e=\mathbf{e}_r \bullet \mathbf{e}_\phi g_r g_e=0$.

By the fundamental hypothesis for the time-retarded theory, the acceleration field $\mathbf{F}_\lambda$ is proportional to the time derivative of $\mathbf{g}_e$. If k is the constant of proportionality, the formula for $\mathbf{F}_\lambda$ is (the formula for the curl can be found in J. D. Jackson's textbook[3])

$$\nabla \times \mathbf{F}_\lambda = \mathbf{e}_\phi k \frac{dg_e}{dt} = \mathbf{e}_\phi \frac{1}{r} \frac{\partial}{\partial r}\left(rF_\lambda\right) \quad . \qquad (2.2.3)$$

Solving for $\partial(rF_\lambda)/\partial r$ and integrating from t=0 to t gives

$$\int_0^t \frac{\partial}{\partial r}\left(rF_\lambda\right) dt = \int_0^t \frac{d}{dt}\left(rF_\lambda\right)\frac{dt}{dr} dt = k\int_0^t r \frac{dg_e}{dt} dt \quad . \qquad (2.2.4)$$

This equation is satisfied for all values of t, if and only if,

$$\int_0^t \left(\frac{d}{dt}\left(rF_\lambda\right)\frac{dt}{dr} - kr\frac{dg_e}{dt}\right) dt = 0 \quad ,$$

which leads to,

$$\frac{d}{dt}\left(rF_\lambda\right) - kr\frac{dg_e}{dt}\frac{dr}{dt} = 0 \quad ,$$

and,

$$F_\lambda(t) = \frac{k}{r}\int_0^t r\,\frac{dg_e}{dt}\frac{dr}{dt}\,dt = \frac{r_E}{r}\int_0^t \frac{r}{r_E}\frac{dg_e}{dt}\frac{u_r}{v_k}\,dt \quad , \qquad (2.2.5)$$

where the induction speed is defined by $v_k=1/k$. The exact value for k is uncertain at this time, but it is known from the NASA flyby data to have the order-of-magnitude value, $k\approx1\times10^{-4}$ s/m ($v_k\approx10$ km/s).(1)

The formula for the classical scalar potential function $\chi_0$ is

$$\chi_0 = -G\iiint \frac{\rho_E dV}{r'} = -\frac{GM_E}{r} = -A\chi_0\frac{r_E}{r} \quad , \qquad (2.2.6)$$

where the formula and numerical value for $A\chi_0$ are

$$A\chi_0 = \frac{GM_E}{r_E} = 6.2595\times10^7\ m^2/s^2 \quad . \qquad (2.2.7)$$

The formula for $\mathbf{g}_r$ becomes

$$\mathbf{g}_r(r) = -\nabla\chi_0(r) = -\nabla\left(\frac{GM_E}{r}\right) = -\mathbf{e}_r\frac{GM_E}{r^2} \quad . \qquad (2.2.8)$$

The formula for the magnitude of the neoclassical vector potential function $\chi_1$ is

$$\chi_1 = G\iiint \frac{1}{c}\frac{dr'}{dt'}\frac{\rho_E dV}{r'} \quad . \qquad (2.2.9)$$

The formula that connects $\mathbf{g}_e$ to $\boldsymbol{\chi}_1$ is

$$\mathbf{g}_e = -\nabla\times\boldsymbol{\chi}_1 \quad . \qquad (2.2.10)$$

Perhaps the easiest way to find the formula for the magnitude of the vector potential $\boldsymbol{\chi}_1$ is to invert the curl operation using the known formula for $\mathbf{g}_e$.(1) Let $\Omega_e$ be the eastward component of the sidereal angular speed for the field point, $\Omega_e\equiv d\phi/dt$, and $\lambda$ is the geocentric latitude for the field point. The Earth's sidereal angular speed is $\Omega_E\equiv d\phi'/dt$. In the $(r,\lambda,\phi,t)$ system, $\mathbf{g}_e=\mathbf{e}_\phi g_e$ and $\boldsymbol{\chi}_1=\mathbf{e}_\lambda\chi_1$. The formula for the signed magnitude of $g_e$ is

$$g_e(r,\lambda) = (-1)A_g\left(\frac{\Omega_e-\Omega_E}{\Omega_E}\right)\cos^2(\lambda)PSr(r) \quad , \qquad (2.2.11)$$

where the coefficient $A_g$ with $c_g=c$ is defined by,

$$A_g \equiv \left(\frac{2r_E\Omega_E}{c}\right)\left(\frac{G}{2}\bar{\rho}_E r_E\right) \quad . \qquad (2.2.12)$$

The power series $PSr(r)$ is defined by,

$$PSr(r) \equiv \left(\frac{r_E}{r}\right)^3 \left(C_0 + C_2\left(\frac{r_E}{r}\right)^2 + C_4\left(\frac{r_E}{r}\right)^4 + C_6\left(\frac{r_E}{r}\right)^6\right) \quad . \tag{2.2.13}$$

Numerical values for the coefficients are

$$C_0 = 0.7018 \;,\; C_2 = 0.2245 \;,\; C_4 = -0.0720 \;,\; C_6 = 0.2671 \;. \tag{2.2.14}$$

Notice that $PSr(r)$ is not a pure inverse-cube function. If the Earth were a homogeneous sphere, i.e., if $\rho_E$ were equal to $\bar{\rho}_E$, the coefficient $C_0$ would equal 1 and $C_2$, $C_4$, and $C_6$ would be zero, in which case $PSr(r)$ would be a pure inverse-cube function.

The formula for the curl operation in spherical coordinates can be found in J. D. Jackson's textbook;[3]

$$\mathbf{e}_\phi \left(\nabla \times (\mathbf{e}_\lambda \chi_1)\right) = \mathbf{e}_\phi \left(\frac{1}{r}\frac{\partial}{\partial r}(r\chi_1)\right) \quad . \tag{2.2.15}$$

Solving for the partial derivative gives

$$\frac{\partial}{\partial r}(r\chi_1) = -rg_e \quad . \tag{2.2.16}$$

Solving this formula for $\chi_1$ and substituting Eq. (2.2.11) for $g_e$ gives

$$\chi_1 = \frac{(-1)}{r}\int_r^\infty rg_e(r,\lambda)dr = A_g r_E\left(\frac{\Omega_e - \Omega_E}{\Omega_E}\right)\cos^2(\lambda)\,\frac{r_E}{r}\int_r^\infty \frac{r}{r_E}\,PSr(r)\,\frac{dr}{r_E} \quad . \tag{2.2.17}$$

Given a value for $r \geq r_E$, the integral over r can be solved easily by using numerical integration, but a power series representation will be useful. Let $PS\chi_1(r)$ be a four-term power series, defined as

$$PS\chi_1(r) = \left(\frac{r_E}{r}\right)^2 \left(C1_0 + C1_2\left(\frac{r_E}{r}\right)^2 + C1_4\left(\frac{r_E}{r}\right)^4 + C1_6\left(\frac{r_E}{r}\right)^6\right) \quad . \tag{2.2.18}$$

By using a least-squares fitting routine, the following coefficients were found to provide an excellent fit of the power series to the integral.

$$C1_0 = 2.9872 \times 10^{-5} \;, \qquad C1_2 = 2.4383 \times 10^{-4} \;,$$
$$C1_4 = -1.5753 \times 10^{-4} \;, \qquad C1_6 = 1.2952 \times 10^{-4} \;. \tag{2.2.19}$$

Now the formula for $\chi_1$ can be rewritten as

$$\chi_1 = A_g r_E\left(\frac{\Omega_e - \Omega_E}{\Omega_E}\right)\cos^2(\lambda)PS\chi_1(r) = A\chi_1\left(\frac{\Omega_e - \Omega_E}{\Omega_E}\right)\cos^2(\lambda)PS\chi_1(r) \quad , \tag{2.2.20}$$

where the formula and numerical value for $A\chi_1$ are

$$A\chi_1 = A_g r_E = 23.1579\, m^2/s^2 \quad . \tag{2.2.21}$$

Let $\boldsymbol{\chi}_2$ be the vector potential for $\mathbf{F}_\lambda$. The formula that connects $\mathbf{F}_\lambda$ to $\boldsymbol{\chi}_2$ is

$$\mathbf{F}_\lambda = \mathbf{e}_\lambda F_\lambda = -\nabla \times \mathbf{X}_2 = -\mathbf{e}_\lambda \left(\nabla \times (\mathbf{e}_\phi \chi_2)\right) = -\mathbf{e}_\lambda \left(\frac{1}{r}\frac{\partial}{\partial r}\left(r\chi_2\right)\right) \quad . \qquad (2.2.22)$$

Solving for the partial derivative gives

$$\frac{\partial}{\partial r}(r\chi_2) = -rF_\lambda \quad . \qquad (2.2.23)$$

Solving this formula for $\chi_2$ and substituting Eq. (2.2.5) for $F_\lambda$ gives

$$\begin{aligned} \chi_2(t) &= \frac{(-1)}{r}\int_{r(0)}^{r(t)} rF_\lambda dr = \frac{(-1)}{r}\int_0^t rF_\lambda \frac{dr}{dt}dt \\ &= \frac{(-1)}{r}\int_0^t r\left(\frac{1}{r}\int_0^t r\frac{dg_e}{dt}\frac{u_r}{v_k}dt\right)u_r dt \\ &= \frac{(-1)}{r}\int_0^t \left(\int_0^t r\frac{dg_e}{dt}\frac{u_r}{v_k}dt\right)u_r dt \quad . \end{aligned} \qquad (2.2.24)$$

The formula for $dg_e/dt$ is given by (cf. Eq. (2.2.11))

$$\frac{dg_e}{dt} = (-1)A_g \frac{d}{dt}\left(\frac{\Omega_e - \Omega_E}{\Omega_E}\cos^2(\lambda)PSr(r)\right) \quad . \qquad (2.2.25)$$

The Earth's rotational surface speed at the equator, call it $v_{Eq}$, provides a convenient reference speed, $v_{Eq}=r_E\Omega_E=464.58$ m/s. The formula for $\chi_2$ can be rewritten as

$$\begin{aligned} \chi_2 &= A_g r_E \frac{v_{Eq}}{v_k}\frac{v_{Eq}}{r}\int_0^t \left(\int_0^t \frac{r}{r_E}\frac{d}{dt}\left(\frac{\Omega_e - \Omega_E}{\Omega_E}\cos^2(\lambda)PSr(r)\right)\frac{u_r}{v_{Eq}}dt\right)\frac{u_r}{v_{Eq}}dt \\ &= A\chi_2 \frac{v_{Eq}}{r}\int_0^t \left(\int_0^t \frac{r}{r_E}\frac{u_r}{v_{Eq}}\frac{d}{dt}\left(\frac{\Omega_e - \Omega_E}{\Omega_E}\cos^2(\lambda)PSr(r)\right)dt\right)\frac{u_r}{v_{Eq}}dt \quad , \end{aligned}$$

so that,

$$\chi_2(t) = A\chi_2 I\chi_2 \quad , \qquad (2.2.26)$$

where the formula and numerical value for $A\chi_2$ are

$$A\chi_2 \equiv A_g r_E \frac{v_{Eq}}{v_k} = A\chi_1 \frac{v_{Eq}}{v_k} \approx 1.1\ m^2/s^2 \quad , \qquad (2.2.27)$$

and the formula for the double integral $I\chi_2$ is

$$I\chi_2(t) \equiv \frac{v_{Eq}}{r}\int_0^t \left(\int_0^t \frac{r}{r_E}\frac{u_r}{v_{Eq}}\frac{d}{dt}\left(\frac{\Omega_e - \Omega_E}{\Omega_E}\cos^2(\lambda)PSr(r)\right)dt\right)\frac{u_r}{v_{Eq}}dt \quad . \qquad (2.2.28)$$

Notice that $\chi_2$ is zero if $u_r=0$ or if $\Omega_e=\Omega_E$.

## 2.3 World Lines and Proper Times

The "world line" for a small mass m is the path of m in "space-time", the space being three dimensional nonrotating and nonaccelerating (inertial) space and the time being one dimensional "coordinated time" for a background sea of hypothetical synchronized clocks. Let the space-time coordinates for m be those for the field-point in the $(r,\lambda,\phi,t)$ system.

Assume the mass m is an ideal clock. Also assume the proper speed $|u|<<c$ and the absolute magnitude of the potentials $|\chi_0|<<c^2$, $|\boldsymbol{\chi}_1|<<c^2$, and $|\boldsymbol{\chi}_2|<<c^2$. Let $d\mathbf{r}$ and $dt$ be elemental changes of path length along the world line. Let $d\tau$ be the elemental change in proper time during the corresponding change $(d\mathbf{r},dt)$. Clocks record proper time. Moving clocks run slow and clocks in a deeper (more negative) gravitational potential run slow. The formula for proper time in the slow-speed weak-field approximation to first-order in $u^2/c^2$ and $\chi/c^2$ is

$$d\tau = \left(1 - \frac{1}{2}\frac{u^2}{c^2} + \frac{\chi_0}{c^2} + \frac{\chi_1}{c^2} + \frac{\chi_2}{c^2}\right)dt \quad . \tag{2.3.1}$$

To solve this formula requires knowing the proper speed $u(t)$ and the signed magnitudes for $\chi_0(t)$, $\chi_1(t)$, and $\chi_2(t)$, at every instant of coordinate time t along the world line.

Let's do a simple thought experiment. Suppose a clock is transported around the Earth at the equator ($\lambda$=0), firstly in the direction of the Earth's rotation (eastward), and secondly oppositely to the Earth's rotation (westward). Suppose $r=r_E$ for both trips. Suppose the trip ground speed is $+v_{Eq}$ for the eastward trip and is $-v_{Eq}$ for the westward trip. Suppose the time recorded by the traveling clock is compared with the time recorded by another (reference) clock that remains "stationary" on the Earth's surface. Suppose the two clocks are at the same place and are synchronized to read the same time (t=0) before each trip around the Earth.

Let $\Delta t$ be the coordinate time interval (sidereal time) required for each trip. A good first approximation for $\Delta t$ is $2\pi r_E/v_{Eq}$=23.93 hours. Let $\Delta\tau_e$ and $\Delta\tau_w$ be the time recorded by the traveling clock during the eastward and westward trips, respectively. Let $\Delta\tau_0$ be the time recorded by the ground reference clock during each trip. Proper time intervals are found by integration of Eq. (2.3.1).

$$\Delta\tau = \int_0^{\Delta t}\left(1 - \frac{1}{2}\frac{u^2}{c^2} + \frac{\chi_0}{c^2} + \frac{\chi_1}{c^2} + \frac{\chi_2}{c^2}\right)dt \quad . \tag{2.3.2}$$

The coordinate speeds, potentials, and proper times are as follows.

Ground reference clock:

$$u = u_\phi = +v_{Eq} = +464.58\,\mathrm{m/s} \quad ,$$

$$\chi_0 = -A\chi_0 \frac{r_E}{r_E} = -6.2595 \times 10^7 \ m^2/s^2 \quad ,$$

$$\chi_1 = A\chi_1 \left( \frac{\Omega_E - \Omega_E}{\Omega_E} \right) \cos^2(0) PS\chi_1(r_E) = 0 \ m^2/s^2 \quad ,$$

$$\chi_2 = A\chi_2 I\chi_2 = 0 \ m^2/s^2 \quad , \quad (u_r{=}0 \text{ and } dg_e/dt{=}0) \quad ,$$

$$\Delta\tau_0 = \left( 1 - \frac{1}{2} \frac{{v_{Eq}}^2}{c^2} - \frac{A\chi_0}{c^2} \right) \Delta t \quad . \tag{2.3.3}$$

Eastward trip:

$$u = u_\phi = +2v_{Eq} = +929.17 \ m/s \quad ,$$

$$\chi_0 = -A\chi_0 \frac{r_E}{r_E} = -6.2595 \times 10^7 \ m^2/s^2 \quad ,$$

$$\chi_1 = A\chi_1 \left( \frac{2\Omega_E - \Omega_E}{\Omega_E} \right) \cos^2(0) PS\chi_1(r_E) = A\chi_1 PS\chi_1(r_E) = +5.6898 \times 10^{-3} \ m^2/s^2 \quad ,$$

$$\chi_2 = A\chi_2 I\chi_2 = 0 \ m^2/s^2 \quad , \quad (u_r{=}0 \text{ and } dg_e/dt{=}0) \quad .$$

The formula for the proper time difference and its numerical value are

$$\Delta\tau_e - \Delta\tau_0 = \left( 1 - 2\frac{{v_{Eq}}^2}{c^2} - \frac{A\chi_0}{c^2} + \frac{A\chi_1 PS\chi_1(r_E)}{c^2} \right) \Delta t - \left( 1 - \frac{1}{2} \frac{{v_{Eq}}^2}{c^2} - \frac{A\chi_0}{c^2} \right) \Delta t$$

$$= \left( -\frac{3}{2} \frac{{v_{Eq}}^2}{c^2} + \frac{A\chi_1 PS\chi_1(r_E)}{c^2} \right) \Delta t = -310.3866 \ ns \quad . \tag{2.3.4}$$

Westward trip:

$$u = u_\phi = 0 \ m/s \quad ,$$

$$\chi_0 = -A\chi_0 \frac{r_E}{r_E} = -6.2595 \times 10^7 \ m^2/s^2 \quad ,$$

$$\chi_1 = A\chi_1 \left( \frac{0 - \Omega_E}{\Omega_E} \right) \cos^2(0) PS\chi_1(r_E) = -A\chi_1 PS\chi_1(r_E) = -5.6898 \times 10^{-3} \ m^2/s^2 \quad ,$$

$$\chi_2 = A\chi_2 I\chi_2 = 0 \ m^2/s^2 \quad , \quad (u_r{=}0 \text{ and } dg_e/dt{=}0) \quad ,$$

$$\Delta\tau_w - \Delta\tau_0 = \left( 1 - \frac{A\chi_0}{c^2} - \frac{A\chi_1 PS\chi_1(r_E)}{c^2} \right) \Delta t - \left( 1 - \frac{1}{2} \frac{{v_{Eq}}^2}{c^2} - \frac{A\chi_0}{c^2} \right) \Delta t$$

$$= \left( \frac{1}{2} \frac{{v_{Eq}}^2}{c^2} - \frac{A\chi_1 PS\chi_1(r_E)}{c^2} \right) \Delta t = +103.4622 \ ns \quad . \tag{2.3.5}$$

Eastward plus three times westward:

$$\Delta\tau_e + 3\Delta\tau_w = \left( -\frac{3}{2} \frac{{v_{Eq}}^2}{c^2} + \frac{A\chi_1 PS\chi_1(r_E)}{c^2} \right) \Delta t + 3\left( \frac{1}{2} \frac{{v_{Eq}}^2}{c^2} - \frac{A\chi_1 PS\chi_1(r_E)}{c^2} \right) \Delta t$$

$$= -2\frac{A\chi_1 PS\chi_1(r_E)}{c^2} \Delta t = -1.0910 \times 10^{-5} \ ns \quad . \tag{2.3.6}$$

This heuristic exercise shows that the traveling clock, relative to the ground reference clock, would lose about 310 nanoseconds during the eastward trip, and would gain about 103 nanoseconds during the westward trip. Notice that the relatively large effect of the scalar potential, $A\chi_0/c^2 \approx 7.0\times10^{-10}$, cancels out, but the effect of the squared speed ratio, $u_\phi^2/c^2 \approx 2.4\times10^{-12}$, and the effect of the vector potential, $A\chi_1 PS\chi_1(r_E)/c^2 \approx 6\times10^{-20}$, do not cancel out. The small effect of the vector potential can be extracted by calculating $\Delta\tau_e + 3\Delta\tau_w$, which for this thought experiment is very small, about $10^{-5}$ ns/day.

Modern precision clocks can detect a time difference as small as about 1 ns/day, but probably cannot detect a time difference of about $10^{-5}$ ns/day. An instrument designed to detect and measure the transverse vector potential utilizing relativistic time dilation will need to employ a method that suppresses the relatively huge effects of $u^2/c^2$ and $A\chi_0/c^2$.

## 3. INSTRUMENT FOR MEASURING THE EARTH'S TIME-RETARDED TRANSVERSE GRAVITATIONAL VECTOR POTENTIAL

### 3.1 Proper Frequency for a Clock Recording Proper Time

What is a physical clock? By definition a clock is a device which records the number of cycles (periods) of an oscillator. If the proper time $\Delta\tau$ for a clock differs from the coordinate time $\Delta t$, the frequency of the clock's oscillator must differ by a commensurate amount.

The instantaneous proper time for a clock, given by Eq. (2.3.1), is

$$d\tau = dt + \left( -\frac{1}{2}\frac{u(t)^2}{c^2} + \frac{\chi_0(t)}{c^2} + \frac{\chi_1(t)}{c^2} + \frac{\chi_2(t)}{c^2} \right) dt \quad . \tag{3.1.1}$$

Integration over a time interval from 0 to t gives the recorded proper time after an interval of coordinate time t.

$$\tau(t) = t + \int_0^t \frac{d\tau}{dt}\, dt = t + \int_0^t \left( -\frac{1}{2}\frac{u(t)^2}{c^2} + \frac{\chi_0(t)}{c^2} + \frac{\chi_1(t)}{c^2} + \frac{\chi_2(t)}{c^2} \right) dt \quad . \tag{3.1.2}$$

The absolute value for the quantity in parenthesis is very small, orders of magnitude less than one.

Let f be the clock-oscillator's frequency (cycles per second or Hertz). Let $f_0$ be the frequency for $d\tau/dt=0$. The formula for the frequency becomes

$$f(t) = f_0 + \int_0^t \frac{df}{dt}\, dt \quad . \tag{3.1.3}$$

Comparing Eq. (3.1.3) with Eq. (3.1.2) indicates that

$$\frac{1}{f_0}\frac{df}{dt} = \left(-\frac{1}{2}\frac{u(t)^2}{c^2} + \frac{\chi_0(t)}{c^2} + \frac{\chi_1(t)}{c^2} + \frac{\chi_2(t)}{c^2}\right) \quad . \qquad (3.1.4)$$

If the clock is losing time, i.e., if $d\tau/dt<0$, the clock-oscillator's relative frequency is less than 1, i.e., $df/dt<f_0$, and vice versa.

The clock's oscillator may be a flywheel, balance wheel, electromagnetic crystal oscillator, optical oscillator, atomic oscillator, or some other kind of (preferably stable) oscillator. To get a very large value for $f_0$, suppose the clock's oscillator is a red-light laser-diode (as found in laser pointers, laser bar-code readers, etc.), for which the optical wavelength is about 700 nm and the intensity usually is about 1 mW. For red light, the approximate optical frequency is

$$f_0 = \frac{c}{700 \times 10^{-9}\ \mathrm{m}} = 4.28 \times 10^{14}\ \mathrm{Hz} \quad . \qquad (3.1.5)$$

The number of photons per second in a 1 mW beam of red light is very large. The photon energy flow divided by Plank's constant times the frequency gives 1 $mJ/hf_0=3.5\times10^{15}$ photons per second. At this intensity there is no need for a statistical analysis of the flow of photons.

## 3.2 Proper Frequencies for Oscillators Circulating in a Two-Arm Rotor

Schematics for three orientations of a two-arm rotor with laser-diode oscillators at the tips are shown in Fig. 1. Let R be the arm length, let $\alpha$ be the angular position of the cross arm, let $\omega\equiv d\alpha/dt$ be the rotor's angular speed, and let v be the tip ground speed, $v\equiv\omega R$. Suppose R=1 m. The Earth's rotational surface speed at the geocentric latitude $\lambda$ is $v_E(\lambda)=r_E\Omega_E\cos(\lambda)$. If $v=v_{Eq}=v_E(0)=464.58$ m/s, the angular speed $\omega=v_{Eq}/R=464.58$ rad/s and the rotor's period $P_{rot}=2\pi/\omega=13.524$ ms. The rps and rpm would be $rps=1/P_{rot}=73.941$ Hz and $rpm=60/P_{rot}=4463$ rounds/minute. At the tips, the speed would be ultrasonic and the "g-value" would be very large, $v_{Eq}^2/R=21968g_E$, where $g_E$ is the acceleration of gravity at the Earth's surface, $g_E=9.825$ m/s$^2$. To safely achieve such high values, the rotor would need to be appropriately designed, constructed from high strength materials, and operated in a vacuum chamber.

Suppose the optical signal from the tip-diodes is conducted through optical fibers to the center of a hollow axel shaft and out through a rotating fiber-optic coupling to a stationary digital electronic interferometer/spectrometer. Suppose that, in addition, a third identical laser-diode oscillator is located at a convenient stationary point on the axis of the rotor to serve as a reference oscillator. Call the three signals: osc1, osc2, and osc0.

When two light beams of the same intensity, the same polarization, and nearly the same frequency are superimposed, the result is a light beam with two frequencies, half the sum of the two initial frequencies and half the difference between the two initial frequencies. The following trig identity shows this effect.

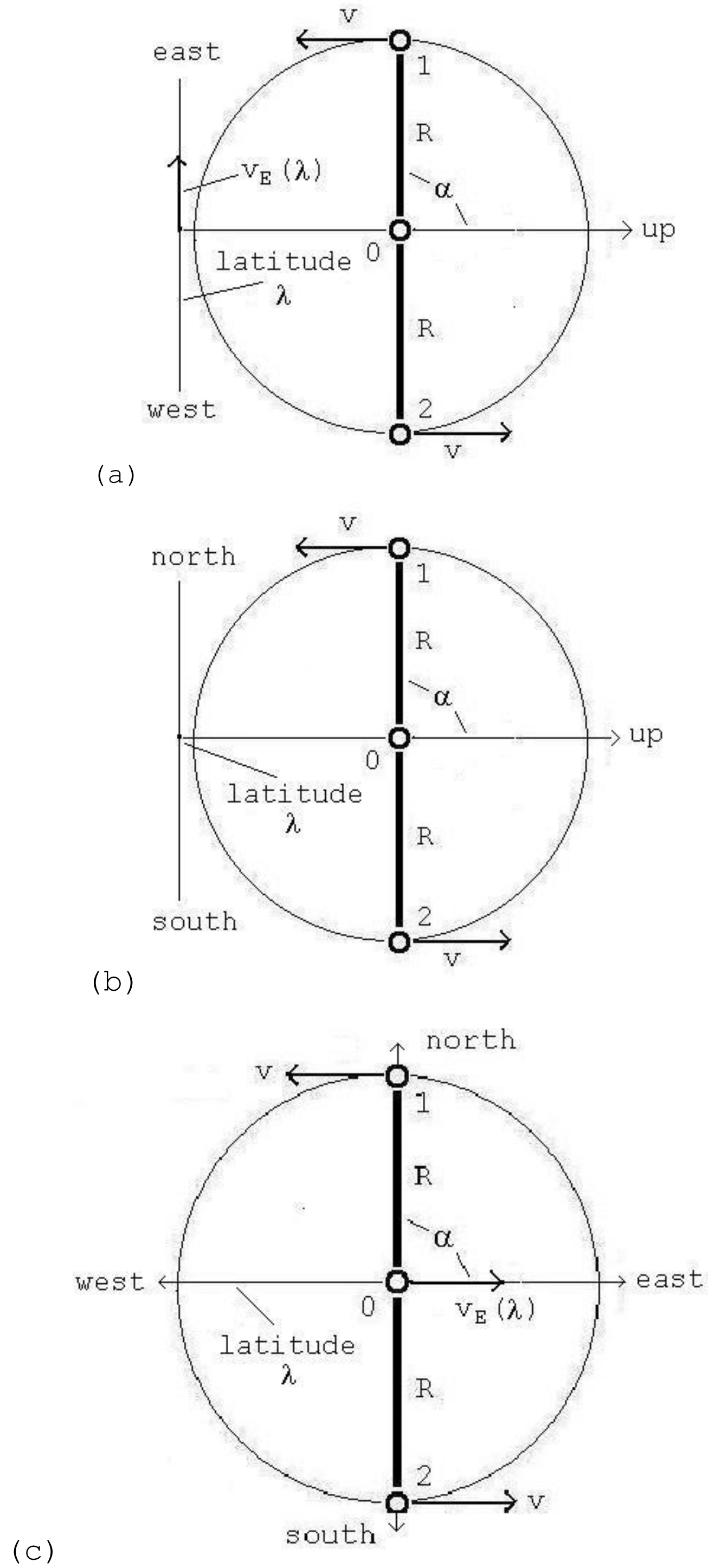

Figure 1. Schematics for a two-arm rotor. (a) Vertical-east-west orientation; the vertical plane of the rotor is moving towards the east with speed $v_E(\lambda)$. (b) Vertical-north-south orientation; the vertical plane of the rotor is moving into the paper with speed $v_E(\lambda)$. (c) Horizontal-east-west orientation; the horizontal plane of the rotor is moving towards the east with speed $v_E(\lambda)$.

$$\cos(\alpha) + \cos(\beta) = 2\cos\left(\frac{1}{2}(\alpha+\beta)\right)\cos\left(\frac{1}{2}(\alpha-\beta)\right) \quad . \tag{3.2.1}$$

The intensity (power in Watts) of a light beam is proportional to the square of the sum. If $\alpha \cong \beta$, the square becomes

$$\begin{aligned}\left(\cos(\alpha) + \cos(\beta)\right)^2 &= 4\cos^2\left(\frac{1}{2}(\alpha-\beta)\right)\cos^2(\alpha) \\ &= (\text{beat signal})(\text{carrier signal}) \quad .\end{aligned} \tag{3.2.2}$$

The factor equaling half the sum is called the "carrier signal", and the factor equaling half the difference is called the "modulation signal" or the "beat signal". If $\alpha$ and $\beta$ are cyclical (harmonic) functions of time, i.e., if $\alpha - \beta \propto \Sigma\cos(n\omega t)$, the beat signal is "frequency modulated" (FM). The frequencies in white light, which contains a continuous spectrum of frequencies, can be identified by using a spectrometer. But for "monochromatic" light, light of a single frequency (color), a spectrometer may not be needed. The FM frequencies for nearly monochromatic light probably can be identified by imposing "amplitude modulation" (AM) on the intensity of the initial laser beams at the frequency of the desired harmonic.

### 3.3 Two-Arm Rotor in the Vertical-East-West Orientation

Consider first the vertical-east-west orientation for the rotor (Fig. 1(a)). Let $r1_{up}$, $r1_n$, and $r1_e$ be the vertical, northward, and eastward components for the radial distance from the Earth's center to the osc1, and let the same notation apply to the other oscillators. For this case,

$$\begin{aligned} r1_{up} &= r_E + R\cos(\omega t) \ , & r1_n &= 0 \ , & r1_e &= R\sin(\omega t) \ , \\ r2_{up} &= r_E - R\cos(\omega t) \ , & r2_n &= 0 \ , & r2_e &= -R\sin(\omega t) \ , \\ r0_{up} &= r_E \ , & r0_n &= 0 \ , & r0_e &= 0 \ . \end{aligned} \tag{3.3.1}$$

Let r1, r2, and r0 be the absolute magnitude for the radial distance from the Earth's center to osc1, osc2, and osc0, respectively. To first order in $R/r_E$,

$$\begin{aligned} r1 &= \left(r1_{up}^2 + r1_n^2 + r1_e^2\right)^{\frac{1}{2}} = r_E\left(1 + \frac{R}{r_E}\cos(\omega t)\right) \ , \\ r2 &= \left(r2_{up}^2 + r2_n^2 + r2_e^2\right)^{\frac{1}{2}} = r_E\left(1 - \frac{R}{r_E}\cos(\omega t)\right) \ , \\ r0 &= \left(r0_{up}^2 + r0_n^2 + r0_e^2\right)^{\frac{1}{2}} = r_E \ . \end{aligned} \tag{3.3.2}$$

Let $u1_{up}$, $u1_n$, and $u1_e$ be the upward component, the northward component, and the eastward component of the proper velocity of osc1, and let the same notation apply to osc2 and osc0.

$$u1_{up} = -v\sin(\omega t)\ ,\qquad u1_n = 0\ ,\qquad u1_e = v_E + v\cos(\omega t)\ ,$$
$$u2_{up} = +v\sin(\omega t)\ ,\qquad u2_n = 0\ ,\qquad u2_e = v_E - v\sin(\omega t)\ ,$$
$$u0_{up} = 0\ ,\qquad u0_n = 0\ ,\qquad u0_e = v_E\ . \tag{3.3.3}$$

The squared magnitude for the velocity is the sum of the squares of the components.

$$u1^2 = \left(u1_{up}^2 + u1_n^2 + u1_e^2\right) = v_E^2\left(1 + \frac{v^2}{v_E^2} + \frac{v}{v_E}\cos(\omega t)\right)\ ,$$
$$u2^2 = \left(u2_{up}^2 + u2_n^2 + u2_e^2\right) = v_E^2\left(1 + \frac{v^2}{v_E^2} - \frac{v}{v_E}\cos(\omega t)\right)\ ,$$
$$u0^2 = v_E^2\ . \tag{3.3.4}$$

The frequencies of the oscillators are reduced by half of the square of the speed ratios. The differences become

$$u1^2 - u0^2 = v_E^2\left(\frac{v^2}{v_E^2} + \frac{v}{v_E}\cos(\omega t)\right) = v^2 + v_{Eq}v\cos(\lambda)\cos(\omega t)\ ,$$
$$u2^2 - u0^2 = v_E^2\left(\frac{v^2}{v_E^2} - \frac{v}{v_E}\cos(\omega t)\right) = v^2 - v_{Eq}v\cos(\lambda)\cos(\omega t)\ ,$$
$$u1^2 - u2^2 = v_E^2\left(2\frac{v}{v_E}\cos(\omega t)\right) = 2v_{E_q}v\cos(\lambda)\cos(\omega t)\ . \tag{3.3.5}$$

Notice that there are no overtones, and that the beat signal for $u1^2$-$u2^2$ is proportional to v.

The frequencies of the oscillators are also changed by the potentials. Let $\chi 01$, $\chi 02$, and $\chi 00$ be the scalar potential values for osc1, osc2, and osc0, respectively. To first order in $R/r_E$, the scalar potentials and the differences reduce to (cf. Eq. (2.2.6))

$$\chi 01 = -\frac{A\chi_0 r_E}{r1} = -A\chi_0\left(1 + \frac{R}{r_E}\cos(\omega t)\right)^{-1} = -A\chi_0\left(1 - \frac{R}{r_E}\cos(\omega t)\right)\ ,$$
$$\chi 02 = -\frac{A\chi_0 r_E}{r2} = -A\chi_0\left(1 - \frac{R}{r_E}\cos(\omega t)\right)^{-1} = -A\chi_0\left(1 + \frac{R}{r_E}\cos(\omega t)\right)\ ,$$
$$\chi 00 = -A\chi_0\frac{r_E}{r0} = -A\chi_0\ ,$$
$$\chi 01 - \chi 00 = A\chi_0\frac{R}{r_E}\cos(\omega t)\ ,$$
$$\chi 02 - \chi 00 = -A\chi_0\frac{R}{r_E}\cos(\omega t)\ ,$$
$$\chi 01 - \chi 02 = 2A\chi_0\frac{R}{r_E}\cos(\omega t)\ . \tag{3.3.6}$$

Notice that the beat signal for the scalar potential $\chi_0$ is independent of v and, up to the first order in $R/r_E$, contains no overtones.

Let $\Omega 1_e$, $\Omega 2_e$, and $\Omega 0_e$ be the eastward component of the angular speed of osc1, osc2, and osc0, respectively.

$$\Omega 1_e = \frac{u1_e}{r_E \cos(\lambda)} = \Omega_E \left(1 + \frac{v}{v_E} \sin(\omega t)\right) ,$$

$$\Omega 2_e = \frac{u2_e}{r_E \cos(\lambda)} = \Omega_E \left(1 - \frac{v}{v_E} \sin(\omega t)\right) ,$$

$$\Omega 0_e = \frac{u0_e}{r_E \cos(\lambda)} = \Omega_E \quad . \qquad (3.3.7)$$

The angular speed ratios become

$$\frac{\Omega 1_e - \Omega_E}{\Omega_E} = + \frac{v}{v_E} \sin(\omega t) ,$$

$$\frac{\Omega 2_e - \Omega_E}{\Omega_E} = - \frac{v}{v_E} \sin(\omega t) ,$$

$$\frac{\Omega 0_e - \Omega_E}{\Omega_E} = 0 \quad . \qquad (3.3.8)$$

The vector potential $\chi_1$ is given by Eq. (2.2.20). Let $\chi 11$ be the $\chi_1$ value for osc1, $\chi 12$ for osc2, and $\chi 10$ for osc0.

$$\chi 11 = A\chi_1 \cos^2(\lambda) \left(\frac{\Omega 1_e - \Omega_E}{\Omega_E}\right) PS\chi_1(r1)$$

$$= A\chi_1 \cos(\lambda) \frac{v}{v_{Eq}} \sin(\omega t) PS\chi_1 \left(r_E \left(1 + \frac{R}{r_E} \cos(\omega t)\right)\right) ,$$

$$\chi 12 = -A\chi_1 \cos(\lambda) \frac{v}{v_{Eq}} \sin(\omega t) PS\chi_1 \left(r_E \left(1 - \frac{R}{r_E} \cos(\omega t)\right)\right) ,$$

$$\chi 10 = 0 \ , \ (\Omega 0_e = \Omega_E) \quad . \qquad (3.3.9)$$

The power series for $\chi_1$ is given by Eq. (2.2.18). To first order in $R/r_E$,

$$PS\chi_1(r1) = \left( \begin{array}{l} C1_0 \left(1 + \frac{R}{r_E} \cos(\omega t)\right)^{-2} + C1_2 \left(1 + \frac{R}{r_E} \cos(\omega t)\right)^{-4} \\ + C1_4 \left(1 + \frac{R}{r_E} \cos(\omega t)\right)^{-6} + C1_6 \left(1 + \frac{R}{r_E} \cos(\omega t)\right)^{-8} \end{array} \right)$$

$$= \left(C1_0 + C1_2 + C1_4 + C1_6 - \left(2C1_0 + 4C1_2 + 6C1_4 + 8C1_6\right) \frac{R}{r_E} \cos(\omega t)\right)$$

$$= C11 \left(1 - \frac{C12}{C11} \frac{R}{r_E} \cos(\omega t)\right) , \qquad (3.3.10)$$

where the sums C11 and C12 , their numerical value, and the ratio are

$$C11 = C1_0 + C1_2 + C1_4 + C1_6 = 2.4570 \times 10^{-4} ,$$

$$C12 = 2C1_0 + 4C1_2 + 6C1_4 + 8C1_6 = 1.1261 \times 10^{-3} \ ,$$
$$\frac{C12}{C11} = 4.5832 \ . \qquad (3.3.11)$$

The formula for $PS\chi_1(r2)$ is similar,

$$PS\chi_1(r2) = \left( C1_0 + C1_2 + C1_4 + C1_6 + \left(2C1_0 + 4C1_2 + 6C1_4 + 8C1_6\right) \frac{R}{r_E} \cos(\omega t) \right)$$
$$= C11 \left( 1 + \frac{C12}{C11} \frac{R}{r_E} \cos(\omega t) \right) \ . \qquad (3.3.12)$$

By using the trig identity $2\sin(\omega t)\cos(\omega t) = \sin(2\omega t)$, the formulas for $\chi11$, $\chi12$, and $\chi10$, and the differences, reduce to

$$\chi11 - \chi10 = \chi11 = A\chi_1 C11 \frac{v}{v_{Eq}} \cos(\lambda) \left( \sin(\omega t) - \frac{1}{2} \frac{C12}{C11} \frac{R}{r_E} \sin(2\omega t) \right) \ ,$$
$$\chi12 - \chi10 = \chi12 = -A\chi_1 C11 \frac{v}{v_{Eq}} \cos(\lambda) \left( \sin(\omega t) + \frac{1}{2} \frac{C12}{C11} \frac{R}{r_E} \sin(2\omega t) \right) \ ,$$
$$\chi11 - \chi12 = 2A\chi_1 C11 \frac{v}{v_{Eq}} \cos(\lambda) \sin(\omega t) \ . \qquad (3.3.13)$$

This shows that the amplitudes for the FM harmonics in the beat signal for $\chi_1$ are proportional to $A\chi_1 v$ (cf. Eq. (2.2.21)). The first overtone for $\chi11-\chi10$ and $\chi12-\chi10$ is proportional to $A\chi_1 v\, C12$. The first overtone for $\chi11-\chi12$ cancels out.

The time derivatives for r1 and r2 are needed to calculate values for the vector potential $\chi_2$.

$$\frac{dr1}{dt} = u1_{up} = -v \sin(\omega t) = -R\omega \sin(\omega t) \ ,$$
$$\frac{dr2}{dt} = u2_{up} = +v \sin(\omega t) = +R\omega \sin(\omega t) \ . \qquad (3.3.14)$$

The formula for $\chi_2$ is given by Eq. (2.2.24). Let $\chi21$ be the value of $\chi_2$ for osc1, and let the same notation apply to osc2 and osc0. By substituting the components for $r_{up}$ of Eq. (3.3.1), the angular speed ratios of Eq. (3.3.8), the values for $u_{up}$ of Eq. (3.3.14), and the formula $v=R\omega$, the formulas for $\chi_2$ become

$$\chi21 = A\chi_2 \cos(\lambda) \frac{v^2}{{v_{Eq}}^2} \frac{v}{r_E} I\chi21 = A\chi_2 \cos(\lambda) \frac{v^2}{{v_{Eq}}^2} \frac{R}{r_E} \omega I\chi21 \ ,$$
$$I\chi21 = \left( 1 - \frac{R}{r_E} \cos(\omega t) \right) \int_0^t I21 \sin(\omega t) dt \ ,$$
$$I21 = \int_0^t \left( 1 + \frac{R}{r_E} \cos(\omega t) \right) \sin(\omega t) \frac{d}{dt} \left( \sin(\omega t) PS r(r1) \right) dt \ ,$$

$$\chi 22 = -A\chi_2 \cos(\lambda) \frac{v^2}{v_{Eq}{}^2} \frac{R}{r_E} \omega I\chi 22 \quad ,$$

$$I\chi 22 = \left(1 + \frac{R}{r_E}\cos(\omega t)\right) \int_0^t I22 \sin(\omega t) dt \quad ,$$

$$I22 = \int_0^t \left(1 - \frac{R}{r_E}\cos(\omega t)\right) \sin(\omega t) \frac{d}{dt}\left(\sin(\omega t) PS\mathrm{r}(r2)\right) dt \quad , \qquad (3.3.15)$$

where I21 and I22 are the first integrals for $I\chi 21$ and $I\chi 22$, respectively.

The formula for $PS\mathrm{r}(r)$ is given by Eq. (2.2.13). By substituting the trig identity $2\sin(\omega t)\cos(\omega t)=\sin(2\omega t)$, and to first order in $R/r_E$, the formula for the argument for the derivative in I21 reduces to

$$\sin(\omega t) PS\mathrm{r}(r1) = \sin(\omega t) \begin{pmatrix} C_0\left(1 + \frac{R}{r_E}\cos(\omega t)\right)^{-3} + C_2\left(1 + \frac{R}{r_E}\cos(\omega t)\right)^{-5} \\ +C_4\left(1 + \frac{R}{r_E}\cos(\omega t)\right)^{-7} + C_6\left(1 + \frac{R}{r_E}\cos(\omega t)\right)^{-9} \end{pmatrix}$$

$$= C1\left(\sin(\omega t) - \frac{1}{2}\frac{C2}{C1}\frac{R}{r_E}\sin(2\omega t)\right) \quad , \qquad (3.3.16)$$

where the sums C1, C2, the ratio, and their numerical values, are

$$C1 = C_0 + C_2 + C_4 + C_6 = 1.1214 \quad ,$$

$$C2 = 3C_0 + 5C_2 + 7C_4 + 9C_6 = 5.1278 \quad ,$$

$$\frac{C2}{C1} = 4.5727 \quad . \qquad (3.3.17)$$

The formula for the argument for the derivative in I22 reduces to

$$\sin(\omega t) PS\mathrm{r}(r2) = C1\left(\sin(\omega t) + \frac{1}{2}\frac{C2}{C1}\sin(2\omega t)\right) \quad . \qquad (3.3.18)$$

Therefore the time derivatives reduce to,

$$\frac{d}{dt}\left(\sin(\omega t) PS\mathrm{r}(r1)\right) = C1\omega\left(\cos(\omega t) - \frac{C2}{C1}\cos(2\omega t)\right)$$

$$= C1\omega\left(\cos(\omega t) - \frac{C2}{C1}\left(2\cos^2(\omega t) - 1\right)\right)$$

$$= C1\omega\left(\frac{C2}{C1} + \cos(\omega t) - 2\frac{C2}{C1}\cos^2(\omega t)\right) \quad ,$$

$$\frac{d}{dt}\left(\sin(\omega t) PS\mathrm{r}(r2)\right) = C1\omega\left(\cos(\omega t) + \frac{C2}{C1}\cos(2\omega t)\right)$$

$$= C1\omega\left(-\frac{C2}{C1} + \cos(\omega t) + 2\frac{C2}{C1}\cos^2(\omega t)\right) \quad . \qquad (3.3.19)$$

Because $R/r_E \ll C2/C1$, the formulas for I21 and I22 reduce to

$$I21 = C1\omega\int_0^t \begin{pmatrix} \frac{C2}{C1}\sin(\omega t) + \sin(\omega t)\cos(\omega t) \\ -2\frac{C2}{C1}\sin(\omega t)\cos^2(\omega t) - 2\frac{C2}{C1}\frac{R}{r_E}\sin(\omega t)\cos^3(\omega t) \end{pmatrix} dt \quad ,$$

$$I22 = C1\omega\int_0^t \begin{pmatrix} -\frac{C2}{C1}\sin(\omega t) + \sin(\omega t)\cos(\omega t) \\ +2\frac{C2}{C1}\sin(\omega t)\cos^2(\omega t) - 2\frac{C2}{C1}\frac{R}{r_E}\sin(\omega t)\cos^3(\omega t) \end{pmatrix} dt \quad . \qquad (3.3.20)$$

The solution for these integrals can be found by applying the following formula from the integral tables.

$$\int_0^t \sin(\omega t)\cos^n(\omega t)dt = \frac{1}{\omega(n+1)}\left(1 - \cos^{(n+1)}(\omega t)\right) \quad , \quad n\neq -1 \quad . \qquad (3.3.21)$$

The solutions are

$$I21 = C1 \begin{pmatrix} \frac{C2}{C1}\left(1 - \cos(\omega t)\right) + \frac{1}{2}\left(1 - \cos^2(\omega t)\right) \\ -2\frac{C2}{C1}\frac{1}{3}\left(1 - \cos^3(\omega t)\right) - 2\frac{C2}{C1}\frac{R}{r_E}\frac{1}{4}\left(1 - \cos^4(\omega t)\right) \end{pmatrix}$$

$$= C1 \begin{pmatrix} \frac{1}{2} + \frac{1}{3}\frac{C2}{C1} - \frac{C2}{C1}\cos(\omega t) - \frac{1}{2}\cos^2(\omega t) \\ +\frac{2}{3}\frac{C2}{C1}\cos^3(\omega t) + \frac{1}{2}\frac{C2}{C1}\frac{R}{r_E}\cos^4(\omega t) \end{pmatrix} \quad ,$$

$$I22 = C1 \begin{pmatrix} \frac{1}{2} - \frac{1}{3}\frac{C2}{C1} + \frac{C2}{C1}\cos(\omega t) - \frac{1}{2}\cos^2(\omega t) \\ -\frac{2}{3}\frac{C2}{C1}\cos^3(\omega t) + \frac{1}{2}\frac{C2}{C1}\frac{R}{r_E}\cos^4(\omega t) \end{pmatrix} \quad . \qquad (3.3.22)$$

By retaining terms to first order in $R/r_E$, and ignoring the constant term, the formulas for $I\chi21$ and $I\chi22$ reduce to

$$I\chi21 = \frac{C1}{\omega} \begin{pmatrix} -\left(\frac{1}{2} + \frac{1}{3}\frac{C2}{C1}\right)\cos(\omega t) + \frac{1}{2}\frac{C2}{C1}\cos^2(\omega t) + \frac{1}{6}\cos^3(\omega t) \\ -\frac{1}{6}\frac{C2}{C1}\cos^4(\omega t) - \frac{1}{10}\frac{C2}{C1}\frac{R}{r_E}\cos^5(\omega t) \end{pmatrix} \quad ,$$

$$I\chi22 = \frac{C1}{\omega} \begin{pmatrix} -\left(\frac{1}{2} - \frac{1}{3}\frac{C2}{C1}\right)\cos(\omega t) - \frac{1}{2}\frac{C2}{C1}\cos^2(\omega t) + \frac{1}{6}\cos^3(\omega t) \\ +\frac{1}{6}\frac{C2}{C1}\cos^4(\omega t) - \frac{1}{10}\frac{C2}{C1}\frac{R}{r_E}\cos^5(\omega t) \end{pmatrix} \quad . \qquad (3.3.23)$$

The trig identities for $\cos^2(\omega t)$, $\cos^3(\omega t)$, $\cos^4(\omega t)$, and $\cos^5(\omega t)$ are

$$\cos^2(\omega t) = \frac{1}{2} + \frac{1}{2}\cos(2\omega t) \quad ,$$

$$\cos^3(\omega t) = \frac{3}{4}\cos(\omega t) + \frac{1}{4}\cos(3\omega t) \quad ,$$

$$\cos^4(\omega t) = \frac{3}{8} + \frac{4}{8}\cos(2\omega t) + \frac{1}{8}\cos(4\omega t) \quad ,$$
$$\cos^5(\omega t) = \frac{10}{16}\cos(\omega t) + \frac{5}{16}\cos(3\omega t) + \frac{1}{16}\cos(5\omega t) \quad . \qquad (3.3.24)$$

By substituting these trig identities into Eq. (3.3.23), the formulas for $I\chi 21$ and $I\chi 22$ reduce to

$$I\chi 21 = \frac{C1}{\omega}\begin{pmatrix} -\left(\frac{3}{8} + \frac{1}{3}\frac{C2}{C1}\right)\cos(\omega t) + \frac{1}{6}\frac{C2}{C1}\cos(2\omega t) \\ + \frac{1}{24}\cos(3\omega t) - \frac{1}{48}\frac{C2}{C1}\cos(4\omega t) - \frac{1}{160}\frac{C2}{C1}\frac{R}{r_E}\cos(5\omega t) \end{pmatrix} \quad ,$$
$$I\chi 22 = \frac{C1}{\omega}\begin{pmatrix} -\left(\frac{3}{8} - \frac{1}{3}\frac{C2}{C1}\right)\cos(\omega t) - \frac{1}{6}\frac{C2}{C1}\cos(2\omega t) \\ + \frac{1}{24}\cos(3\omega t) + \frac{1}{48}\frac{C2}{C1}\cos(4\omega t) - \frac{1}{160}\frac{C2}{C1}\frac{R}{r_E}\cos(5\omega t) \end{pmatrix} \quad . \qquad (3.3.25)$$

Because $\chi 20 = 0$, and to first order in $R/r_E$, the beat signals reduce to

$$\chi 21 - \chi 20 = \chi 21 = A\chi_2 \cos(\lambda)\frac{v^2}{{v_{Eq}}^2}\frac{R}{r_E}\omega I\chi 21$$
$$= A\chi_2 \cos(\lambda)\frac{v^2}{{v_{Eq}}^2}\frac{R}{r_E} C1 \begin{pmatrix} -\left(\frac{3}{8} + \frac{1}{3}\frac{C2}{C1}\right)\cos(\omega t) + \frac{1}{6}\frac{C2}{C1}\cos(2\omega t) \\ + \frac{1}{24}\cos(3\omega t) - \frac{1}{48}\frac{C2}{C1}\cos(4\omega t) \end{pmatrix} \quad ,$$
$$\chi 22 - \chi 20 = \chi 22 = -A\chi_2 \cos(\lambda)\frac{v^2}{{v_{Eq}}^2}\frac{R}{r_E}\omega I\chi 22$$
$$= A\chi_2 \cos(\lambda)\frac{v^2}{{v_{Eq}}^2}\frac{R}{r_E} C1 \begin{pmatrix} \left(\frac{3}{8} - \frac{1}{3}\frac{C2}{C1}\right)\cos(\omega t) + \frac{1}{6}\frac{C2}{C1}\cos(2\omega t) \\ - \frac{1}{24}\cos(3\omega t) - \frac{1}{48}\frac{C2}{C1}\cos(4\omega t) \end{pmatrix} \quad ,$$
$$\chi 21 - \chi 22 = A\chi_2 \cos(\lambda)\frac{v^2}{{v_{Eq}}^2}\frac{R}{r_E} C1 \left(-\frac{3}{4}\cos(\omega t) + \frac{1}{12}\cos(3\omega t)\right) \quad . \qquad (3.3.26)$$

This shows that the beat signals for $\chi_2$ are proportional to $v^2$. The beat signal for $\chi 21 - \chi 20$ contains a rich array of overtones, overtones at $2\omega t$, $3\omega t$, and $4\omega t$. But the overtones at $2\omega t$ and $4\omega t$ cancel out of the beat signal for $\chi 21 - \chi 22$.

Let $\delta u^2$, $\delta\chi_0$, $\delta\chi_1$, and $\delta\chi_2$ be the time dilation effects of the FM harmonic amplitudes for $u^2$, $\chi_0$, $\chi_1$, and $\chi_2$ in the beat signals for the vertical-east-west orientation.

$$\delta u^2 = \frac{1}{2} f_0 \frac{\left(u1^2 - u0^2\right)}{c^2} = \frac{1}{2}\frac{f_0 v_E v}{c^2}\left(\cos(\omega t)\right) \quad ,$$
$$= \frac{1}{2} f_0 \frac{\left(u1^2 - u2^2\right)}{c^2} = \frac{f_0 v_E v}{c^2}\left(\cos(\omega t)\right) \quad ,$$

Table I. Order-of-magnitude estimates of the FM harmonic amplitudes for a two-arm rotor in the vertical-east-west orientation for osc1-osc0 with $\lambda=0$ and $v=v_{Eq}$. The harmonic frequencies are $\omega/2\pi=73.9$ Hz, $2\omega/2\pi=148.0$ Hz, $3\omega/2\pi=221.8$ Hz, and $4\omega/2\pi=295.6$ Hz.

| harmonic | $\omega t$ | $2\omega t$ | $3\omega t$ | $4\omega t$ |
|---|---|---|---|---|
| $\delta u^2$ | $5.1\times10^{2}$ Hz | 0 | 0 | 0 |
| $\delta\chi_0$ | $4.7\times10^{-2}$ Hz | 0 | 0 | 0 |
| $\delta\chi_1$ | $2.7\times10^{-5}$ Hz | $9.8\times10^{-12}$ Hz | 0 | 0 |
| $\delta\chi_2$ | $1.7\times10^{-9}$ Hz | $6.9\times10^{-10}$ Hz | $3.8\times10^{-11}$ Hz | $8.6\times10^{-11}$ Hz |

$$\delta\chi_0 = f_0\frac{(\chi01-\chi00)}{c^2} = \frac{f_0A\chi_0}{c^2}\frac{R}{r_E}(\cos(\omega t)) ,$$

$$= f_0\frac{(\chi01-\chi02)}{c^2} = 2\frac{f_0A\chi_0}{c^2}\frac{R}{r_E}(\cos(\omega t)) ,$$

$$\delta\chi_1 = f_0\frac{(\chi11-\chi10)}{c^2} = \frac{f_0A\chi_1}{c^2}\frac{v}{v_{Eq}}\cos(\lambda)\begin{cases} C11(\sin(\omega t)) \\ \frac{1}{2}\frac{R}{r_E}C12(\sin(2\omega t)) \end{cases} ,$$

$$= f_0\frac{(\chi11-\chi12)}{c^2} = 2\frac{f_0A\chi_1}{c^2}\frac{v}{v_{Eq}}\cos(\lambda)C11(\sin(\omega t)) ,$$

$$\delta\chi_2 = f_0\frac{(\chi21-\chi20)}{c^2} = \frac{f_0A\chi_2}{c^2}\cos(\lambda)\frac{v^2}{v_{Eq}^2}\frac{R}{r_E}\begin{cases} \left(\frac{3}{8}C1+\frac{1}{3}C2\right)(\cos(\omega t)) \\ \frac{1}{6}C2(\cos(2\omega t)) \\ \frac{1}{24}C1(\cos(3\omega t)) \\ \frac{1}{48}C2(\cos(4\omega t)) \end{cases} ,$$

$$= f_0\frac{(\chi21-\chi22)}{c^2} = \frac{f_0A\chi_2}{c^2}\cos(\lambda)\frac{v^2}{v_{Eq}^2}\frac{R}{r_E}\begin{cases} \frac{3}{4}C1\cos(\omega t) \\ \frac{1}{12}C1\cos(3\omega t) \end{cases} . \quad (3.3.27)$$

The expected order-of-magnitude FM harmonic amplitudes for the vertical-east-west orientation at the equator ($\lambda=0$) with $v=v_{Eq}$ are listed in Table I. This table shows that the amplitude for the fundamental is dominated by $\delta u^2$, but the amplitudes at $2\omega t$, $3\omega t$, and $4\omega t$ are dominated by $\delta\chi_2$. By allowing sufficient rotor operating time, say more than 1 hour=3600 s, and by using sufficient amplification, say more than $10^6$, these overtones probably will be detectable.

### 3.4 Two-Arm Rotor in the Vertical-North-South Orientation

Next consider the vertical-north-south orientation for the rotor (Fig. 1(b)). For this case, $\Omega_e=\Omega_E$ for all three oscillators, which means the vector potentials equal zero for all three oscillators. Also the squared speeds cancel out in the osc1-osc2 beat signal. This orientation provides a convenient reference for the time dilation effect of only the scalar potential.

Again let $r1_{up}$, $r1_n$, and $r1_e$ be the vertical, northward, and eastward components for the radial distance from the Earth's center to the osc1, and let the same notation apply to the other oscillators. For this case,

$$r1_{up} = r_E + R\cos(\omega t) \ , \quad r1_n = R\sin(\omega t) \ , \quad r1_e = 0 \ ,$$
$$r2_{up} = r_E - R\cos(\omega t) \ , \quad r2_n = -R\sin(\omega t) \ , \quad r2_e = 0 \ ,$$
$$r0_{up} = r_E \ , \quad r0_n = 0 \ , \quad r0_e = 0 \ . \qquad (3.4.1)$$

Let r1, r2, and r0 be the absolute magnitude for the radial distance to osc1, osc2, and osc0, respectively. To first order in $R/r_E$,

$$r1 = \left(r1_{up}{}^2 + r1_n{}^2 + r1_e{}^2\right)^{\frac{1}{2}} = r_E\left(1 + \frac{R}{r_E}\cos(\omega t)\right) \ ,$$
$$r2 = \left(r2_{up}{}^2 + r2_n{}^2 + r2_e{}^2\right)^{\frac{1}{2}} = r_E\left(1 - \frac{R}{r_E}\cos(\omega t)\right) \ ,$$
$$r0 = \left(r0_{up}{}^2 + r0_n{}^2 + r0_e{}^2\right)^{\frac{1}{2}} = r_E \ . \qquad (3.4.2)$$

Again let $u1_{up}$, $u1_n$, and $u1_e$ be the upward, northward, and eastward components of the proper velocity for osc1, and let the same notation apply to osc2 and osc0.

$$u1_{up} = -v\sin(\omega t) \ , \quad u1_n = v\cos(\omega t) \ , \quad u1_e = v_E \ ,$$
$$u2_{up} = v\sin(\omega t) \ , \quad u2_n = -v\cos(\omega t) \ , \quad u2_e = v_E \ ,$$
$$u0_{up} = 0 \ , \quad u0_n = 0 \ , \quad u0_e = v_E \ . \qquad (3.4.3)$$

The formulas for $u^2$ are

$$u1^2 = v_E{}^2\left(1 + \frac{v^2}{v_E{}^2}\right) \ ,$$
$$u2^2 = v_E{}^2\left(1 + \frac{v^2}{v_E{}^2}\right) \ ,$$
$$u0^2 = v_E{}^2 \ . \qquad (3.4.4)$$

The differences become

$$u1^2 - u0^2 = v^2 \ ,$$
$$u2^2 - u0^2 = v^2 \ ,$$
$$u1^2 - u2^2 = 0 \ . \qquad (3.4.5)$$

Again let $\chi01$, $\chi02$, and $\chi00$ be the scalar potential values for osc1, osc2, and osc0, respectively. To first order in $R/r_E$, the scalar potentials and the differences become

$$\chi 01 = -\frac{A\chi_0 r_E}{r1} = -A\chi_0\left(1 - \frac{R}{r_E}\cos(\omega t)\right) ,$$

$$\chi 02 = -\frac{A\chi_0 r_E}{r2} = -A\chi_0\left(1 + \frac{R}{r_E}\cos(\omega t)\right) ,$$

$$\chi 00 = -A\chi_0 \frac{r_E}{r0} = -A\chi_0 ,$$

$$\chi 01 - \chi 00 = A\chi_0 \frac{R}{r_E}\cos(\omega t) ,$$

$$\chi 02 - \chi 00 = -A\chi_0 \frac{R}{r_E}\cos(\omega t) ,$$

$$\chi 01 - \chi 02 = 2A\chi_0 \frac{R}{r_E}\cos(\omega t) . \quad (3.4.6)$$

The numerical value for $\delta\chi_0$ is

$$\delta\chi_0 = 2\frac{f_0 A\chi_0}{c^2}\frac{R}{r_E}\cos(\omega t) = 9.4 \times 10^{-2}\ \mathrm{Hz}\left(\cos(\omega t)\right) . \quad (3.4.7)$$

This shows that the vertical-north-south orientation provides a good way to check on the effect of the scalar potential acting alone.

### 3.5 Two-Arm Rotor in the Horizontal-East-West Orientation

Finally consider the horizontal-east-west orientation for the rotor (Fig. 1(c)). This orientation provides a convenient check for time dilation effects without the effect of the scalar potential.

Again let $r1_{up}$, $r1_n$, and $r1_e$ be the vertical, northward, and eastward components for the radial distance from the Earth's center to the osc1, and let the same notation apply to the other oscillators. For this case,

$$r1_{up} = r_E , \qquad r1_n = R\sin(\omega t) , \qquad r1_e = R\cos(\omega t) ,$$
$$r2_{up} = r_E , \qquad r2_n = -R\sin(\omega t) , \qquad r2_e = -R\cos(\omega t) ,$$
$$r0_{up} = r_E , \qquad r0_n = 0 , \qquad r0_e = 0 . \quad (3.5.1)$$

Let r1, r2, and r0 be the absolute magnitude for the radial distance to osc1, osc2, and osc0, respectively. To order $R/r_E$,

$$r1 = \left(r1_{up}^{\ 2} + r1_n^{\ 2} + r1_e^{\ 2}\right)^{\frac{1}{2}} = \left(r_E^{\ 2} + R^2\sin^2(\omega t) + R^2\cos^2(\omega t)\right)^{\frac{1}{2}} = r_E ,$$

$$r2 = \left(r2_{up}^{\ 2} + r2_n^{\ 2} + r2_e^{\ 2}\right)^{\frac{1}{2}} = r_E ,$$

$$r0 = \left(r0_{up}^{\ 2} + r0_n^{\ 2} + r0_e^{\ 2}\right)^{\frac{1}{2}} = r_E . \quad (3.5.2)$$

Again let $u1_{up}$, $u1_n$, and $u1_e$ be the upward component, the northward component, and the eastward component of the proper velocity of osc1, and let the same notation apply to osc2 and osc0.

$$u1_{up} = 0 \;, \qquad u1_n = v\cos(\omega t) \;, \qquad u1_e = v_E - v\sin(\omega t) \;,$$
$$u2_{up} = 0 \;, \qquad u2_n = -v\cos(\omega t) \;, \qquad u2_e = v_E + v\sin(\omega t) \;,$$
$$u0_{up} = 0 \;, \qquad u0_n = 0 \;, \qquad u0_e = v_E \;. \tag{3.5.3}$$

The squared magnitude for the velocity is the sum of the squares of the components.

$$u1^2 = \left(v^2\cos^2(\omega t) + \left(v_E - v\sin(\omega t)\right)^2\right) = v_E^{\,2}\left(1 + \frac{v^2}{v_E^{\,2}} - 2\frac{v}{v_E}\sin(\omega t)\right) \;,$$
$$u2^2 = \left(v^2\cos^2(\omega t) + \left(v_E + v\sin(\omega t)\right)^2\right) = v_E^{\,2}\left(1 + \frac{v^2}{v_E^{\,2}} + 2\frac{v}{v_E}\sin(\omega t)\right) \;,$$
$$u0^2 = v_E^{\,2} \;. \tag{3.5.4}$$

The differences become

$$u1^2 - u0^2 = v^2 - 2v_E v\sin(\omega t) \;,$$
$$u2^2 - u0^2 = v^2 + 2v_E v\sin(\omega t) \;,$$
$$u1^2 - u2^2 = -4v_E v\sin(\omega t) \;. \tag{3.5.5}$$

Again let $\chi01$, $\chi02$, and $\chi00$ be the scalar potential values for osc1, osc2, and osc0, respectively. To order $R/r_E$, the scalar potentials and the differences become

$$\chi01 = -\frac{A\chi_0 r_E}{r1} = -A\chi_0\left(1 + \frac{1}{2}\frac{R^2}{r_E^{\,2}}\right)^{-1} = -A\chi_0 \;,$$
$$\chi02 = -\frac{A\chi_0 r_E}{r2} = -A\chi_0\left(1 + \frac{1}{2}\frac{R^2}{r_E^{\,2}}\right)^{-1} = -A\chi_0 \;,$$
$$\chi00 = -A\chi_0\frac{r_E}{r0} = -A\chi_0 \;,$$
$$\chi01 - \chi00 = 0 \;,$$
$$\chi02 - \chi00 = 0 \;,$$
$$\chi01 - \chi02 = 0 \;. \tag{3.5.6}$$

Let $\Omega1_e$, $\Omega2_e$, and $\Omega0_e$ be the eastward component of the angular speed of osc1, osc2, and osc0, respectively.

$$\Omega1_e = \frac{u1_e}{r_E\cos(\lambda)} = \frac{v_E}{r_E\cos(\lambda)}\left(1 - \frac{v}{v_E}\sin(\omega t)\right) = \Omega_E\left(1 - \frac{v}{v_E}\sin(\omega t)\right) \;,$$
$$\Omega2_e = \frac{u2_e}{r_E\cos(\lambda)} = \Omega_E\left(1 + \frac{v}{v_E}\sin(\omega t)\right) \;,$$
$$\Omega0_e = \frac{u0_e}{r_E\cos(\lambda)} = \frac{v_E}{r_E\cos(\lambda)} = \Omega_E \;. \tag{3.5.7}$$

The angular speed ratios become

$$\frac{\Omega 1_e - \Omega_E}{\Omega_E} = + \frac{v}{v_E} \sin(\omega t) \ ,$$
$$\frac{\Omega 2_e - \Omega_E}{\Omega_E} = - \frac{v}{v_E} \sin(\omega t) \ ,$$
$$\frac{\Omega 0_e - \Omega_E}{\Omega_E} = 0 \ . \qquad (3.5.8)$$

The vector potential $\chi_1$ is given by Eq. (2.2.20). Let $\chi 11$ be the $\chi_1$ value for osc1, $\chi 12$ for osc2, and $\chi 10$ for osc0.

$$\chi 11 = A\chi_1 \cos^2(\lambda) \left( \frac{\Omega 1_e - \Omega_E}{\Omega_E} \right) PS\chi_1(r1)$$
$$= A\chi_1 \cos^2(\lambda) \frac{v}{v_{Eq} \cos(\lambda)} \sin(\omega t) PS\chi_1 \left( r_E \left( 1 + \frac{R}{r_E} \cos(\omega t) \right) \right)$$
$$= A\chi_1 \cos(\lambda) \frac{v}{v_{Eq}} \sin(\omega t) C11 \left( 1 - \frac{C12}{C11} \frac{R}{r_E} \cos(\omega t) \right) \ ,$$
$$\chi 12 = -A\chi_1 \cos(\lambda) \frac{v}{v_{Eq}} \sin(\omega t) C11 \left( 1 + \frac{C12}{C11} \frac{R}{r_E} \cos(\omega t) \right) \ ,$$
$$\chi 10 = 0 \ . \qquad (3.5.9)$$

By using the trig identity for $\sin(\omega t)\cos(\omega t)$, the formulas for $\chi 11$, $\chi 12$, and the difference, can be rewritten as

$$\chi 11 - \chi 10 = \chi 11 = A\chi_1 \frac{v}{v_{Eq}} \cos(\lambda) \left( C11 \sin(\omega t) - \frac{1}{2} C12 \frac{R}{r_E} \sin(2\omega t) \right) \ ,$$
$$\chi 12 - \chi 10 = \chi 12 = -A\chi_1 \frac{v}{v_{Eq}} \cos(\lambda) \left( C11 \sin(\omega t) + \frac{1}{2} C12 \frac{R}{r_E} \sin(2\omega t) \right) \ ,$$
$$\chi 11 - \chi 12 = 2A\chi_1 \frac{v}{v_{Eq}} C11 \cos(\lambda) \sin(\omega t) \ . \qquad (3.5.10)$$

Because $u1_{up}$, $u2_{up}$, and $u0_{up}$ equal zero, all the values for $\chi_2$ are zero.

Again let $\delta u^2$, $\delta\chi_0$, $\delta\chi_1$, and $\delta\chi_2$ be the time dilation effects of the FM harmonic amplitudes for $u^2$, $\chi_0$, $\chi_1$, and $\chi_2$ in the beat signals for the horizontal-east-west orientation.

$$\delta u^2 = \frac{1}{2} f_0 \frac{\left( u1^2 - u0^2 \right)}{c^2} = \frac{f_0 v_E v}{c^2} \left( \cos(\omega t) \right) \ ,$$
$$= \frac{1}{2} f_0 \frac{\left( u1^2 - u2^2 \right)}{c^2} = 2 \frac{f_0 v_E v}{c^2} \left( \cos(\omega t) \right) \ ,$$
$$\delta\chi_0 = f_0 \frac{(\chi 01 - \chi 00)}{c^2} = 0 \ ,$$
$$= f_0 \frac{(\chi 01 - \chi 02)}{c^2} = 0 \ ,$$

$$\delta\chi_1 = f_0 \frac{(\chi 11 - \chi 10)}{c^2} = \frac{f_0 A\chi_1}{c^2} \frac{v}{v_{Eq}} \cos(\lambda) \begin{cases} C11\left(\sin(\omega t)\right) \\ \frac{1}{2}\frac{R}{r_E} C12\left(\sin(2\omega t)\right) \end{cases} ,$$

$$= f_0 \frac{|\chi 11 - \chi 12|}{c^2} = 2\frac{f_0 A\chi_1}{c^2} \frac{v}{v_{Eq}} \cos(\lambda) C11\left(\sin(\omega t)\right) ,$$

$$\delta\chi_2 = f_0 \frac{(\chi 21 - \chi 20)}{c^2} = 0 ,$$

$$= f_0 \frac{(\chi 21 - \chi 22)}{c^2} = 0 . \quad (3.5.11)$$

These results show that the horizontal-east-west orientation provides a good way to check for the effect on the first overtone for the vector potential $\chi_1$ acting alone.

**CONCLUSIONS AND RECOMMENDATIONS**

This feasibility study indicates that a high-speed two-arm rotor can be designed to provide precision measurements for the induction constant k, the corresponding induction speed $v_k=1/k$, and the radial mass density coefficient sums C1, C2, C11, and C12. The instrument can be made portable by installing it in a semi-tractor-trailer system. Then it could be used to measure the harmonic amplitudes at different latitudes and different altitudes.

Of course, many engineering details need to be worked out to perfect this basic design. That takes only time, money, and desire.

**ACKNOWLEDGEMENTS**

I thank Patrick L. Ivers for reviewing the original manuscript and suggesting improvements.